\documentclass[twocolumn,showpacs,preprintnumbers,amsmath,amssymb]{revtex4}

\def\undersim#1{\setbox9\hbox{${#1}$}{#1}\kern-\wd9\lower
    2.5pt \hbox{\lower\dp9\hbox to \wd9{\hss $_\sim$\hss}}}
\usepackage{amssymb}
\usepackage{amsmath}
\usepackage{graphicx}
\usepackage{color}

\def\undersim#1{\setbox9\hbox{${#1}$}{#1}\kern-\wd9\lower
    2.5pt \hbox{\lower\dp9\hbox to \wd9{\hss $_\sim$\hss}}}

\def\mK{{\mathbf K}}

\def\mq{{\mathbf q}}
\def\mk{{\mathbf k}}

\begin{document}

\title{Squeezing effect on three-dimensional Hanbury Brown-Twiss radii\footnote{Supported by the National Natural Science Foundation of China
under Grant No. 11905085; Guangdong Basic and Applied Basic Research Foundation under Grant No. 2022A1515110392.}}

\author{Yong Zhang$^{1}$}
\author{Peng Ru$^{2}$}
\affiliation{\small$^1$School of Mathematics and Physics,
Jiangsu University of Technology, Changzhou, Jiangsu 213001, China\\
$^2$School of Materials and New Energy, South China Normal University, Shanwei 516699, China}


\begin{abstract}
This paper examines the impacts of the squeezing effect caused by the particle's
in-medium mass modification on the three-dimensional Hanbury Brown-Twiss (HBT) radii.
An analysis is conducted on how the squeezing effect impacts the three-dimensional
HBT radii of $\phi\phi$, $D^0$$D^0$, and $K^+$$K^+$. The squeezing effect suppresses the impacts
of transverse flow on the transverse source distribution and broadens the
space-time rapidity distribution of the particle-emitting source,
leading to an increase in the HBT radii, notably in out and longitudinal direction.
This phenomenon becomes more significant for higher transverse pair momentum,
resulting in a non-monotonic decrease in the HBT radii with increasing transverse
pair momentum. The impact of the squeezing effect on the HBT radii is more pronounced
for $D^0$$D^0$ than for $\phi\phi$. Furthermore, this effect is also more significant for $\phi$$\phi$ than for $K^+$$K^+$.
The findings presented in this paper could offer
fresh perspectives on investigating the squeezing effect.

Keywords: Squeezing effect; in-medium mass modification; HBT radii; $\phi\phi$; $D^0$$D^0$; $K^+$$K^+$.

\end{abstract}

\maketitle

\section{Introduction}
The Hanbury Brown-Twiss (HBT) correlation, commonly referred to as Bose-Einstein correlation, has become an important observable for investigating the spatial and temporal structure of the particle-emitting sources created in relativistic heavy-ion collisions \cite{Gyu79,Wongbook,Wie99,Wei00,Csorgo02,Lisa05}.
HBT radii are key parameters in HBT analyses that provide insights into the spatial and temporal dimensions of the sources.

Because of their interactions with the source medium, the mass of the particles in the medium may vary from their mass in a vacuum. This variation can result in a squeezing effect, leading to an experimental observable known as squeezed back-to-back correlation (SBBC) between bosons and antibosons \cite{AsaCso96,AsaCsoGyu99,Padula06,DudPad10,Zhang15a,YZHANG_CPC15,Zhang-EPJC16,AGY17,XuZhang19,Zhang2024}. For sources that have a broad temporal distribution, the SBBC may be completely suppressed \cite{DudPad10,Zhang-EPJC16}. Consequently, even with the squeezing effect, SBBC is unable to provide feedback for sources with broad temporal distribution.
The squeezing effect influences the HBT correlation by reducing the impact of flow \cite{DudPad10,Zhang2024},
leading to a non-flow behavior in the one-dimensional HBT radii. Even for sources with broad temporal distributions, this non-flow behavior persists in the HBT radii \cite{Zhang2024}. Thus, the non-flow behavior of the HBT radii offers a new way to study squeezing effects beyond SBBC.
The previous research on the influences of the squeezing effect on HBT was based on the spherically symmetric Gaussian
expanding source \cite{DudPad10,Zhang2024}. Actually, the sources formed in relativistic heavy-ion collisions are not spherically symmetric.
In addition, one-dimensional HBT radii can not provide detailed information for each dimension.
It is necessary to further study the impact of the squeezing effect on three-dimensional HBT radii.

In this paper, a locally thermalized cylinder expansion source \cite{s1,s2} is used to study the influences of the squeezing effect on three-dimensional HBT radii.
The squeezing effect on the three-dimensional HBT radii of $\phi$$\phi$ and $D^0$$D^0$ are shown.
Since the impacts of the squeezing effect on the HBT results are more pronounced for the bosons with large masses \cite{Zhang2024}.
In addition, due to their electrically neutral nature, $\phi$ and $D^0$ remain uninfluenced by the Coulomb effect, making them ideal probes for investigating the squeezing effect compared to charged bosons.

Currently, there are no HBT experimental measurements for $\phi$$\phi$ and $D^0$$D^0$. The majority of experiments are concentrated on HBT measurement and analysis involving two pions and two kaons \cite{phbt1,phbt2,phbt3,phbt4,phbt5,phbt6,khbt1,khbt2,khbt3,khbt4,khbt5,khbt6}.
Kaons have a relatively larger mass, which may better show the squeezing effect on their HBT radii compared to pions.
Thus, the impacts of squeezing effect on the three-dimensional HBT radii of $K^+$$K^+$ are also analyzed in this paper. In the calculations, the Coulomb effect is not considered.

The squeezing effect affects the HBT radii by suppressing the influence of transverse flow on the transverse source distribution and increasing the width of the space-time rapidity distribution of the particle-emitting source , ultimately leading to an increase in HBT radii, especially for large transverse pair momentum $K_T$.
Consequently, the HBT radii no longer monotonically decreasing with increasing $K_T$. This phenomenon is referred to the non-flow behavior of the HBT radii \cite{Zhang2024}.
This strange behavior of HBT radii may offer new insights into the study of squeezing effects.
The squeezing effect has a more significant impact on $R_o$ and $R_l$ than $R_s$.
The squeezing effect has a greater influence on HBT radii for the boson with large mass and for low freeze-out temperatures.

The $\phi$, $D$, and kaon mesons serve as excellent probes for investigating the quark-gluon plasma (QGP) generated during
relativistic heavy-ion collisions due to their inclusion of either a strange or a charm quark, known to experience
the complete evolution of the QGP resulting from these collisions.
A great deal of attention has been triggered by the latest investigations into experimental data related to the
$\phi$, $D$, and kaon mesons \cite{{STAR-PRL21,STAR-PRC20,STAR-PRC19D,CMS-PRL18D,CMS-PLB18D,
ALICE-JHEP22,ALICE-EPJC20,ALICE-JHEP18D,
ALICE-JHEP16D,ALICE-JHEP15D,ALICE-PRC14D,ALICE-PRL13D,ALICE-EPJC23,ALICE-PRC22,ALICE-EPJC21,ALICE-EPJC18p,STAR-PRL16p,
STAR-PRC16p,PHENIX-PRC16p,ALICE-PRC15p,PHENIX-PRC11p,STAR-PLB09p,STAR-PRC09p,
STAR-PRL07p,PHENIX-PRL07p,PHENIX-PRC05p,NA50-PLB03p,STAR-PRC02p,NA50-PLB00p,kaon-e1,kaon-e2,kaon-e3,kaon-e4,kaon-e5,kaon-e6}}.
In addition, it was anticipated that the interactions with the medium would induce modifications to the masses of $\phi$, $D$, and kaon mesons within the particle-emitting sources produced during relativistic heavy-ion collisions \cite{Dm,dm2,dm3,dm4,pm,pm1,pm2,kaonm1,kaonm2,kaonm3}.
Consequently, the analysis in this paper is meaningful in relativistic heavy-ion collisions.

The rest of the paper is organized as follows:
In Section II, the formulas of the HBT correlation function with a squeezing effect for cylinder expansion sources are presented. Additionally, the formulas of the three-dimensional HBT radii based on Gaussian form are shown in this section.
Section III examines the impacts of the squeezing effect on the distributions of sources for the $\phi$, $D^0$, and $K^+$ mesons.
It also analyzes how the squeezing effect influences the three-dimensional HBT radii of $\phi$$\phi$, $D^0$$D^0$,
and $K^+$$K^+$.
Finally, Section IV concludes the paper with a summary and discussion.

\section{Formulas}
The HBT correlation function of two identical bosons was defined as \cite{Gyu79,Wongbook,Wie99,Wei00,Csorgo02,Lisa05}

\begin{eqnarray}
C(\mq,\mK) =1+\frac{|\int d^{4}rS(r,K)e^{iq\cdot r}|^{2}}{\int d^{4}rS(r,k_1)d^{4}rS(r,k_2)}.
\end{eqnarray}
$S(r,k)$ denotes the emission function. The momenta of the two bosons are denoted by
$k_1$ and $k_2$, respectively. Additionally, $\mq = \mk_1-\mk_2$ and $\mK = (\mk_1+\mk_2)/2$ are the relative and pair momentum of two identical bosons.
For cylinder expansion sources, the emission function can be represented as \cite{s1,s2}
\begin{eqnarray}\label{emission}
&&\hspace*{-12mm}S(r,k) = M_T\cosh(\eta-Y)f(r,k)\nonumber\\
&&\hspace*{2mm}\times\exp(-\frac{r_T^2}{2R_g^2}-\frac{\eta^2}{2(\delta\eta)^2})\delta(\tau-\tau_0).
\end{eqnarray}
Here, $M_T = \sqrt{m^2+k_T^2}$ is the transverse mass, where $m$ and $k_T = \sqrt{k_x^2+k_y^2}$ are the mass and the transverse momentum of boson in a vacuum, respectively.
$Y$ is the rapidity of the boson. $r_T = \sqrt{x^2+y^2}$ and $\eta = \frac{1}{2}\ln[(t+z)/(t-z)]$ are the transverse coordinate and the space-time rapidity,
respectively. $R_g$ and $\delta\eta$ are the parameters that govern the width of the transverse distribution and the width of the space-time rapidity distribution, respectively.
$\tau = \sqrt{t^2-z^2}$ is the proper time. In this paper, it is assumed that the bosons freeze-out at a constant proper time $\tau_0$.
$f(r,k)$ is the momentum distribution of the bosons at the freeze-out point $r$.
When the squeezing effect is not taken into account, the $f(r,k)$ for boson is \cite{s1,s2}
\begin{equation}\label{f0}
f(r,k) = \frac{1}{\exp[k^{\mu}u_{\mu}(r)/T]-1}.
\end{equation}
$k^{\mu}=(E_{\mk}=\sqrt{\mk^2+m^2},{\mk})$ is the four-momentum of the boson. $T$ is the freeze-out temperature of the boson. $u_{\mu}(r)$ is the four-velocity,
which was assumed as \cite{s1,s2}
\begin{eqnarray}
u_{\mu}(r) = (\cosh\eta\cosh\eta_T,\sinh\eta_T \vec{e}_T,\sinh\eta\cosh\eta_T),
\end{eqnarray}
Here, $\vec{e}_T = (x/r_T,y/r_T)$. The transverse flow rapidity, denoted as $\eta_T$, is defined as \cite{s2}
\begin{equation}
\begin{cases}
\eta_T= \eta_{Tmax}\frac{r_T}{R_G}  \,\,\,\,\,\,\,\,\,\, r_T<R_G,\\
\eta_T= \eta_{Tmax}              \,\,\,\,\,\,\,\,\,\,\,\,\,\,\,\,\,\,  r_T\geq R_G.
\end{cases}
\end{equation}
The value of $\eta_{Tmax}$ is calculated by $\eta_{Tmax} = \frac{1}{2}\ln(\frac{1+\beta}{1-\beta})$, where $\beta$ is a parameter.

When the squeezing effect is considered, the $f(r,k)$ becomes \cite{Gyu79,Wongbook,Wie99,Wei00,Csorgo02,Lisa05,AsaCso96,AsaCsoGyu99,Padula06,DudPad10,Zhang2024}

\begin{equation}\label{f}
f(r,k) = |c_{\mk'}|^2\,n_{\mk'}+\,|s_{-\mk'}|^2\,(\,n_{-\mk'}+1),
\end{equation}
\begin{equation}\label{csk}
c_{\mk'}=\cosh[\,r_{\mk'}\,], \,\,\,s_{\mk'}=\sinh[\,r_{\mk'}\,],
\end{equation}
\begin{eqnarray}
r_{\mk'}=\frac{1}{2} \log \left[E_{\mk'}/\varepsilon_{\mk'}\right],
\end{eqnarray}
\begin{eqnarray}
E_{\mk'}(r)=\sqrt{\mk'^2(r)+m^2}=k^{\mu} u_{\mu}(r),
\end{eqnarray}
\begin{eqnarray}
&&\hspace*{-7mm}\varepsilon_{\mk'}(r)=\sqrt{\mk'^2(r)+m_*^2}\nonumber\\
&&\hspace*{3.mm}=\sqrt{[E_{\mk'}(r)]^2-m^2+m_*^2},
\end{eqnarray}
\begin{equation}\label{BZ}
n_{\mk'}=\frac{1}{\exp(\varepsilon_{\mk'}(r)/T)-1}.
\end{equation}
Here, $\mk'$ denotes the local-frame momentum corresponding to $\mk$, while $m_*$ refers to
the mass of the boson in the source medium.
When $m_* =$ $ m$, it means there is no change in mass in the medium, causing
$f(r,k)$ in Eq. \ref{f} to equal $f(r,k)$ in Eq. \ref{f0}. This condition of $m_* =$ $ m$ also implies that the squeezing effect is not
considered in the calculations.

The Gaussian form of the HBT correlation function can be expressed as \cite{s2}:
\begin{equation}
C(\mq,\mK) = 1+\lambda\exp[-q^2_oR^2_o(\mK)-q^2_sR^2_s(\mK)-q^2_lR^2_l(\mK)],
\end{equation}
Here, $\lambda$ represents the coherence parameter.
$q_l$ is the spatial component of $\mq$ in the beam direction (``longitudinal" direction), while $q_o$ and $q_s$ are
the spatial components of $\mq$ parallel to the transverse components $\mK_T$ of $\mK$ (``out" direction),
perpendicular to the transverse components $\mK_T$ (``side" direction), respectively.
$R^2_o$, $R^2_s$ and $R^2_l$ were expressed as \cite{s1,s2}
\begin{eqnarray}
	R_o^2 = \langle(r_o-\beta_o t)^2\rangle - \langle r_o-\beta_o t\rangle^2,
\end{eqnarray}
\begin{eqnarray}
	R_s^2 = \langle r_s^2\rangle,
\end{eqnarray}
\begin{eqnarray}
	R_l^2 = \langle(r_l-\beta_l t)^2\rangle - \langle r_l-\beta_l t\rangle^2,
\end{eqnarray}
\begin{eqnarray}
\beta_i = \frac{2K_i}{E_1+E_2} \approx \frac{K_i}{E_K}.
\end{eqnarray}
Here, the average notation is defined as
\begin{eqnarray}
\langle\zeta\rangle = \frac{\int d^{4}r\zeta S(r,K)}{\int d^{4}rS(r,K)}.
\end{eqnarray}

\section{Results}
In this section, the impacts of the squeezing effect on three-dimensional HBT radii of $\phi$$\phi$, $D^0$$D^0$,
and $K^+$$K^+$ are shown.
In the calculations, the parameters $R_G$ and $\delta\eta$ are considered to be 6 fm and 3.0 \cite{s2}, respectively, representing the transverse width and the space-time rapidity width of the source. The parameter $\tau_0$ is taken to be 10 fm/c \cite{s2}. The transverse expansion velocity of the source is determined by the parameter $\beta$ and is set to 0, 0.3, and 0.5.

The masses of the particles $\phi$, $D^0$, and $K^+$ in a vacuum, represented by $m$, are known to be 1.01946 GeV, 1.86484 GeV, and 0.49368 GeV \cite{PDG24}, respectively.
The mass modification in medium is symbolized by $\delta m$, where $\delta m = m-m_* $.
If $\delta m = 0 $, this indicates that the squeezing effect is not considered in the calculations.
In the pionic medium, it is anticipated that the masses of the particles $\phi$, $D^0$, and $K^+$ will be decreased \cite{pm,Dm}. Thus, $\delta m$ is taken as greater than 0 in this paper.
The freeze-out temperature for the $\phi$ meson is considered to be 0.14 GeV \cite{Padula06,Zhang2024}, with a comparative analysis also made at 0.15 GeV.
The freeze-out temperature for the $D^0$ meson is set to 0.14 GeV and 0.15 GeV.
And the freeze-out temperature for the $K^+$ meson is similarly set to 0.14 GeV and 0.15 GeV.
The rapidity $Y$ is considered within the range of $(-1,1)$, except in the longitudinally co-moving system (LCMS) shown in Figs. \ref{roslp}, \ref{rosld} and \ref{roslk}. In the LCMS, $Y$ = 0 \cite{Wie99,s1}.

\subsection{$\phi$$\phi$}

\begin{figure}[htbp]
\vspace*{0mm}
\includegraphics[scale=0.66]{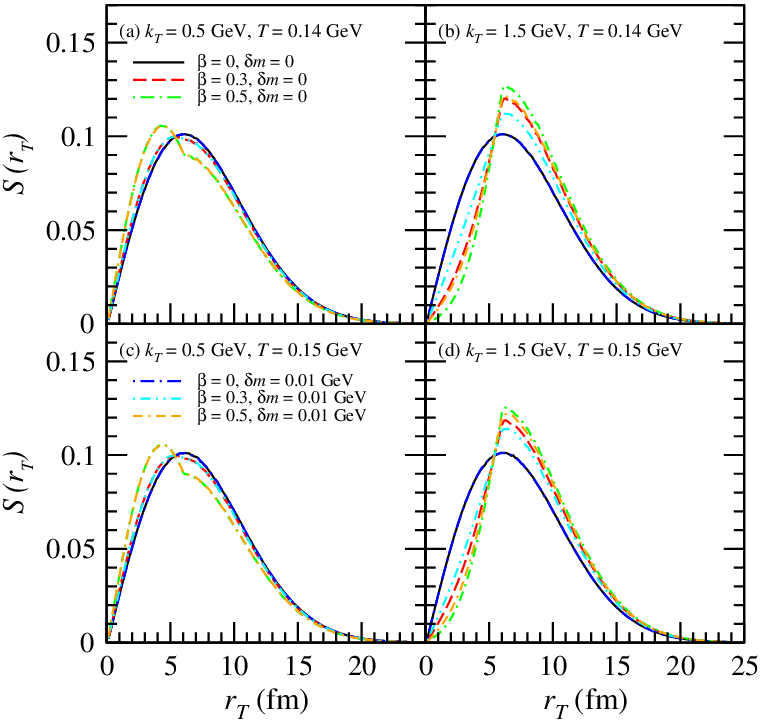}
\vspace*{0mm}
\caption{(Color online) Normalized distributions of transverse coordinate of the $\phi$ emission sources for $k_T$ = 0.5 GeV and 1.5 GeV.
Here, $T$ = 0.14 GeV in plots (a) and (b), while $T$ = 0.15 GeV in plots (c) and (d).  }
\label{psr}
\end{figure}

\begin{figure}[htbp]
\vspace*{0mm}
\includegraphics[scale=0.66]{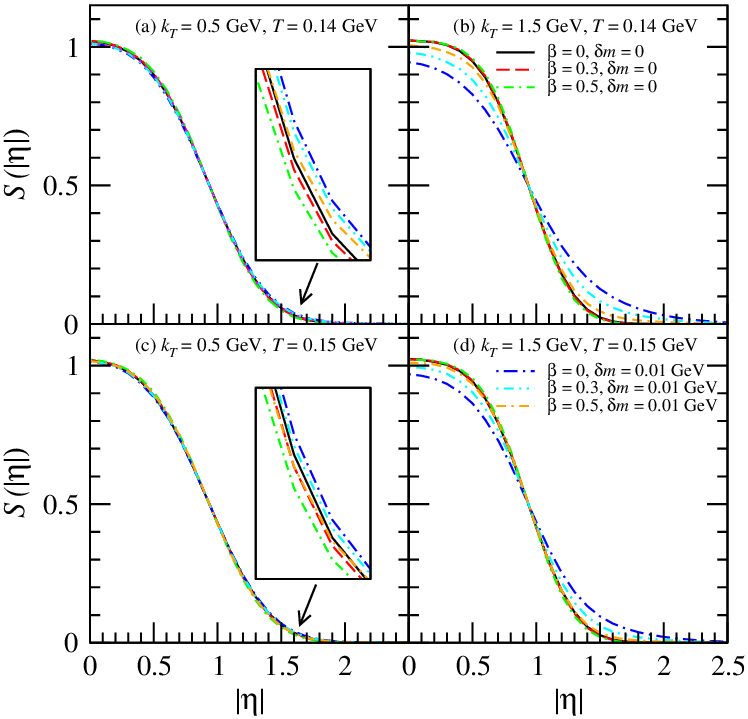}
\vspace*{0mm}
\caption{(Color online) Normalized distributions of $|\eta|$ of the $\phi$ emission sources for $k_T$ = 0.5 GeV and 1.5 GeV. }
\label{psz}
\end{figure}

\begin{figure}[htbp]
\vspace*{0mm}
\includegraphics[scale=0.82]{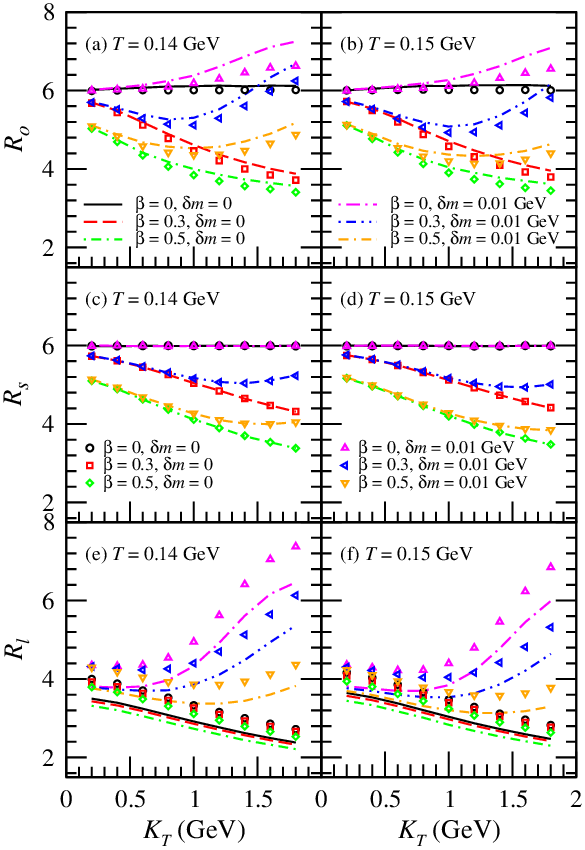}
\vspace*{0mm}
\caption{(Color online) HBT radii $R_o$, $R_s$ and $R_l$ of $\phi$$\phi$ with respect to transverse pair momentum $K_T$. The plots illustrate the results of $R_o$ in (a) and (b),
$R_s$ in (c) and (d), $R_l$ in (e) and (f). The lines are results for $Y$ within the range of $(-1,1)$, while the symbols denote the results calculated in the LCMS. }
\label{roslp}
\end{figure}

In Fig. \ref{psr}, the normalized distributions of transverse coordinate of the $\phi$ emission sources are shown for $k_T$ values of 0.5 GeV and 1.5 GeV.
Here, $T$ = 0.14 GeV in plots (a) and (b), while $T$ = 0.15 GeV in plots (c) and (d).
When $\beta = 0$, it indicates the absence of any velocity expansion in the transverse plane, and the transverse source distributions for $k_T$ = 1.5 GeV
being identical to those for $k_T$ = 0.5 GeV.
When the squeezing effect is not considered ($\delta m = 0 $), the transverse flow leads to a shift in the transverse source distributions for $k_T$ = 0.5 GeV towards smaller transverse coordinate $r_T$, but leads to a shift in the transverse source distributions for $k_T$ = 1.5 GeV towards larger transverse coordinate $r_T$.
This occurs because particles with low transverse momentum are more likely to be produced in greater quantities at lower transverse velocities, while locations with higher transverse velocities tend to generate more high transverse momentum particles. In the model discussed in this paper, the transverse flow increases as the transverse coordinate $r_T$ increases.
With the increase of transverse velocity, this phenomenon becomes more pronounced.
At $T = 0.14$ GeV, this phenomenon exhibits a slightly more significant presence compared to $T = 0.15$ GeV.
The mass of $\phi$ meson is expected to be reduced by around 0.01 GeV in the pionic medium \cite{pm}.
When considering the squeezing effect, the $\delta m$ for $\phi$ meson is taken as 0.01 GeV in Figs. \ref{psr}-\ref{roslp}.
For $\beta = 0$, the squeezing effect does not affect the transverse source distribution.
However, for $\beta > 0$, the squeezing effect reduces the influence of transverse flow on the transverse source distribution at $k_T$ = 1.5 GeV
while having no impact at $k_T$ = 0.5 GeV. This phenomenon is more pronounced for $T$ = 0.14 GeV.

In Fig. \ref{psz}, the normalized distributions of $|\eta|$ of the $\phi$ emission sources for $k_T$ = 0.5 GeV and 1.5 GeV are shown. Here, $\eta$ is the space-time rapidity.
For $\delta m = 0 $, the transverse flow slightly reduces the space-time rapidity distribution of the sources.
The squeezing effect leads to the widening of the source space-time rapidity distribution, which is more obvious for $k_T$ = 1.5 GeV.
This phenomenon is more pronounced for lower transverse flow or smaller freeze-out temperatures.

In the model discussed in this paper, $t = \tau\cosh\eta$ and $z = \tau\sinh\eta$. Therefore, by reducing the width of the source space-time rapidity distribution,
the transverse flow reduces the width of the temporal distribution of and the longitudinal distribution.
However, the squeezing effect leads to the widening of the source space-time rapidity distribution, which results in an expansion of the temporal distribution of and the longitudinal distribution of the source.

In Fig. \ref{roslp}, HBT radii $R_o$, $R_s$ and $R_l$ of $\phi$$\phi$ with respect to transverse pair momentum $K_T$ are shown. The plots illustrate the results of $R_o$ in (a) and (b), $R_s$ in (c) and (d), $R_l$ in (e) and (f). The lines are results for $Y$ within the range of $(-1,1)$, while the symbols denote the results calculated in the LCMS. In this paragraph, the results for $Y$ in the range of $(-1,1)$ are discussed first, followed by a discussion of the differences between these results and those obtained in the LCMS.
The value of $R_o$ is related to the temporal and the transverse spatial distribution of the source.
When the squeezing effect is not considered ($\delta m = 0 $), the transverse flow causes a reduction in transverse spatial distribution width, especially for large $K_T$
(see Fig. \ref{psr}). Furthermore, the transverse flow slightly diminishes the width of the source space-time rapidity distribution (see Fig. \ref{psz}), leading to a decrease in the width of the source temporal distribution.
Therefore, under the influence of the transverse flow, $R_o$ decreases as $K_T$ increases for $\beta > 0$.
The squeezing effect reduces the influence of transverse flow on the transverse source distribution for large $k_T$ (see Fig. \ref{psr} (b) and (d))
while having no impact for small $k_T$ (see Fig. \ref{psr} (a) and (c)). Moreover, the squeezing effect results in the widening of the source space-time rapidity distribution, consequently expanding the temporal distribution of the source, especially for large $k_T$ (see Fig. \ref{psz}). As a result, the squeezing effect leads to an increase in $R_o$,
which is more pronounced at high $K_T$ values.
The value of $R_s$ is only related to the transverse spatial distribution of the source.
When $\beta = 0$, the squeezing effect has no impact on the transverse source distribution (see Fig. \ref{psr}), thus not influencing $R_s$ either.
For $\beta > 0$, the squeezing effect leads to an increase in $R_s$, which is more pronounced at high $K_T$ values. This is due to the impacts of the squeezing effect on the transverse source distribution.
The value of $R_l$ is related to the temporal and the longitudinal spatial distribution of the source.
The squeezing effect leads to the widening of the source space-time rapidity distribution, which results in an expansion of the temporal distribution of and the longitudinal distribution of the source. These impacts are more pronounced for large $k_T$. As a result, the squeezing effect leads to an increase in $R_l$,
which is more pronounced at high $K_T$ values.
The squeezing effect has a more significant impact on $R_o$ and $R_l$ than $R_s$.
The squeezing effect has a slightly greater impact on $R_o$, $R_s$ and $R_l$ at $T=0.14$ GeV compared to $T=0.15$ GeV.
The phenomenon of $R_o$, $R_s$ and $R_l$ no longer monotonically decreasing with increasing $K_T$ due to the squeezing effect being more pronounced at high $K_T$ is referred to the non-flow behavior of the HBT radii \cite{Zhang2024}. $R_o$ calculated in the LCMS is slightly lower than that calculated for $Y$ in the range of $(-1,1)$.
Conversely, $R_s$ calculated in the LCMS is the same as that calculated for $Y$ in the range of $(-1,1)$. However, $R_l$ calculated in the LCMS is greater than that calculated for $Y$ in the range of $(-1,1)$.
The impact of the squeezing effect on the HBT radii in the LCMS is similar to its impact on HBT radii for $Y$ in the range of $(-1,1)$.
When accounting for the squeezing effect, the difference between the values of $R_o$ and $R_l$ calculated in the LCMS and the values of $R_o$ and $R_l$ for $Y$ within the range of $(-1,1)$ is greater than the difference observed without the squeezing effect.

\begin{figure}[htbp]
\vspace*{0mm}
\includegraphics[scale=0.82]{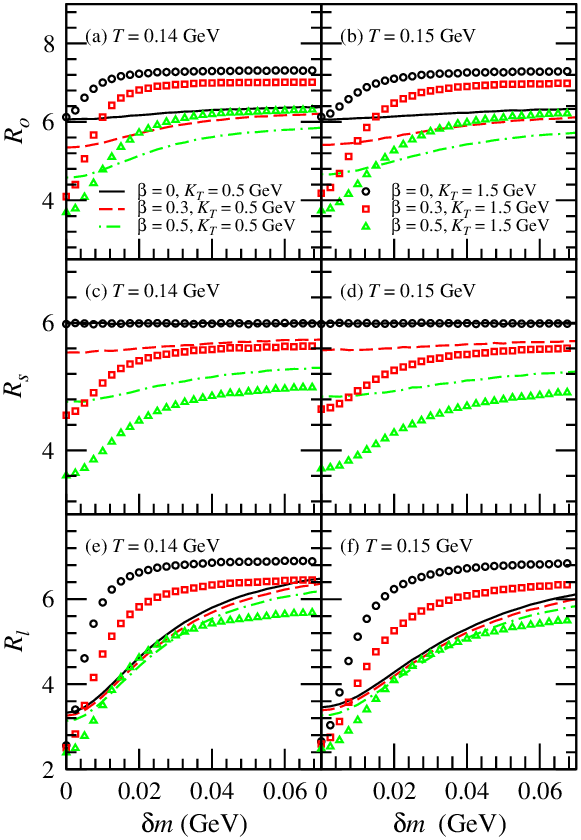}
\vspace*{0mm}
\caption{(Color online) HBT radii $R_o$, $R_s$ and $R_l$ of $\phi$$\phi$ with respect to in-medium mass modification $\delta m$ for $K_T$ = 0.5 GeV and 1.5 GeV. The plots illustrate the results of $R_o$ in (a) and (b), $R_s$ in (c) and (d), $R_l$ in (e) and (f). }
\label{roslpdm}
\end{figure}

In Fig. \ref{roslpdm}, HBT radii $R_o$, $R_s$ and $R_l$ of $\phi$$\phi$ with respect to in-medium mass modification $\delta m$ for $K_T$ = 0.5 GeV and 1.5 GeV are shown.
When $\beta = 0$, the squeezing effect does not affect $R_s$, which remains constant as $\delta m$ increases.
However, in other instances, as $\delta m$ increases, the value of $R_o$, $R_s$ and $R_l$
first increase before plateauing, indicating a decreasing slope in the relationship between $R_o$, $R_s$ and $R_l$ and $\delta m$.
Ultimately, $R_o$, $R_s$ and $R_l$ stabilize with minimal alterations as $\delta m$ further increases.
When $\delta m <$ 0.02 GeV, The slopes of $R_o$, $R_s $, and $R_l $ as a function of $\delta m$ are significantly larger at $K_T$ = 1.5 GeV than at $K_T$ = 0.5 GeV,
and they are slightly larger at $T$ = 0.14 GeV than at $T$ = 0.15 GeV.

\subsection{$D^0$$D^0$}
\begin{figure}[htbp]
\vspace*{0mm}
\includegraphics[scale=0.66]{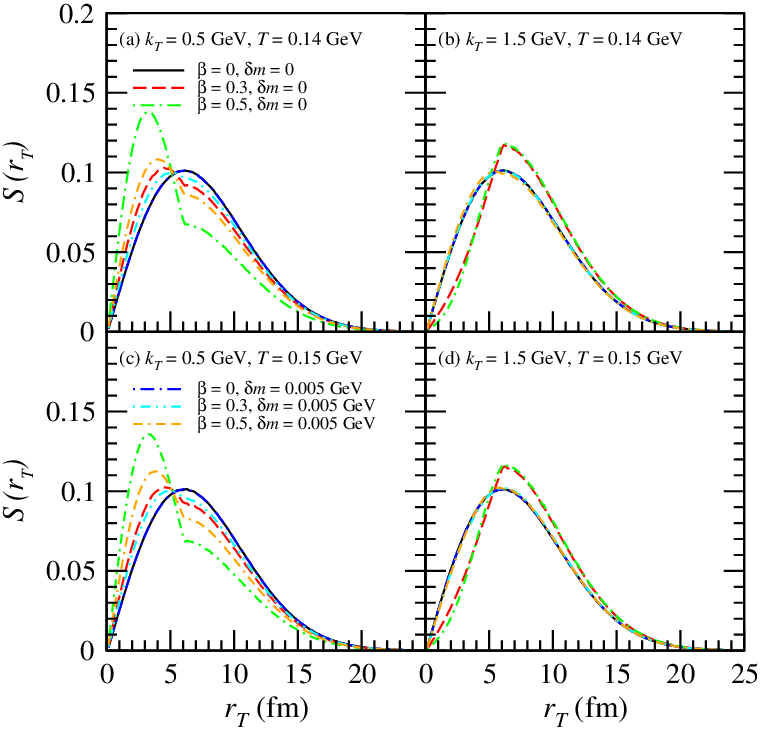}
\vspace*{0mm}
\caption{(Color online) Normalized distributions of transverse coordinate of the $D^0$ emission sources for $k_T$ = 0.5 GeV and 1.5 GeV.
Here, $T$ = 0.14 GeV in plots (a) and (b), while $T$ = 0.15 GeV in plots (c) and (d).  }
\label{psrd}
\end{figure}

\begin{figure}[htbp]
\vspace*{0mm}
\includegraphics[scale=0.66]{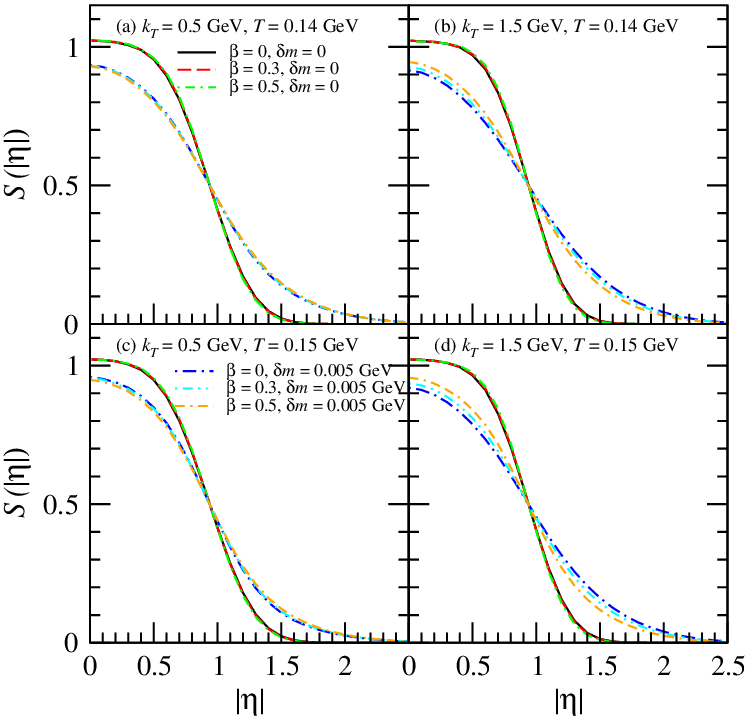}
\vspace*{0mm}
\caption{(Color online) Normalized distributions of $|\eta|$ of the $D^0$ emission sources for $k_T$ = 0.5 GeV and 1.5 GeV. }
\label{pszd}
\end{figure}

\begin{figure}[htbp]
\vspace*{0mm}
\includegraphics[scale=0.82]{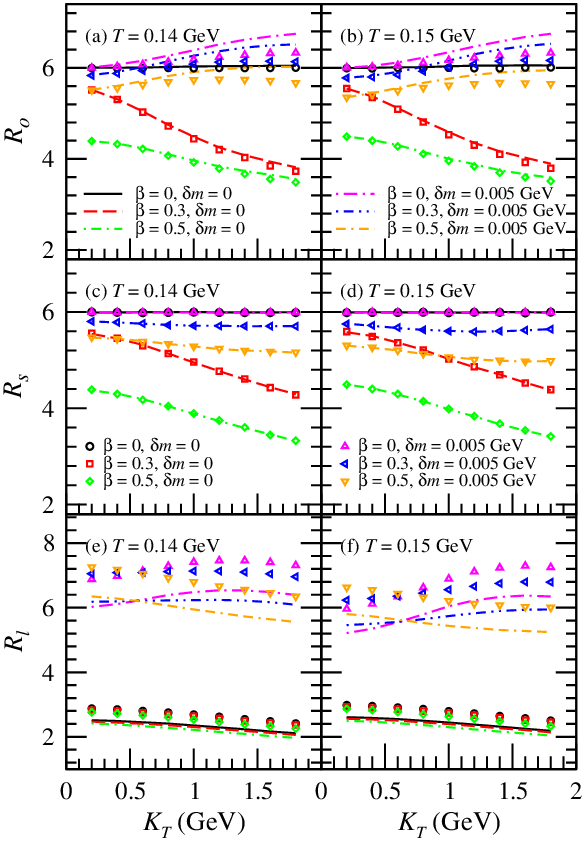}
\vspace*{0mm}
\caption{(Color online) HBT radii $R_o$, $R_s$ and $R_l$ of $D^0$$D^0$ with respect to transverse pair momentum $K_T$. The plots illustrate the results of $R_o$ in (a) and (b),
$R_s$ in (c) and (d), $R_l$ in (e) and (f). The lines are results for $Y$ within the range of $(-1,1)$, while the symbols denote the results calculated in the LCMS. }
\label{rosld}
\end{figure}

In Fig. \ref{psrd}, the normalized distributions of transverse coordinate of the $D^0$ emission sources for $k_T$ = 0.5 GeV and 1.5 GeV are presented.
Here, $T$ = 0.14 GeV in plots (a) and (b), while $T$ = 0.15 GeV in plots (c) and (d).
When $\beta = 0$, it indicates the absence of any velocity expansion in the transverse plane, and the transverse source distributions for $k_T$ = 1.5 GeV
being identical to those for $k_T$ = 0.5 GeV.
When the squeezing effect is not considered ($\delta m = 0 $), the transverse flow leads to a shift in the transverse source distributions for $k_T$ = 0.5 GeV towards smaller transverse coordinate $r_T$, but leads to a shift in the transverse source distributions for $k_T$ = 1.5 GeV towards larger transverse coordinate $r_T$.
The mass of $D^0$ meson is expected to be reduced by around 0.005 GeV in the pionic medium \cite{Dm}.
When considering the squeezing effect, the $\delta m$ for $D^0$ meson is taken as 0.005 GeV in Figs. \ref{psrd}-\ref{rosld}.
For $\beta = 0$, the squeezing effect does not affect the transverse source distribution.
However, for $\beta > 0$, the squeezing effect reduces the influence of transverse flow on the transverse source distribution, which is more pronounced for $k_T$ = 1.5 GeV.
This phenomenon is slightly more pronounced for $T$ = 0.14 GeV.
The squeezing effect affects the transverse source distribution of $D^0$  more notably than that of $\phi$ meson.

In Fig. \ref{pszd}, the normalized distributions of $|\eta|$ of the $D^0$ emission sources for $k_T$ = 0.5 GeV and 1.5 GeV are shown.
The squeezing effect results in the widening of the distribution of $|\eta|$ of the $D^0$ emission sources, which is more pronounced for $k_T$ = 1.5 GeV.
This phenomenon is slightly more pronounced for $T$ = 0.14 GeV.

In Fig. \ref{rosld}, the HBT radii $R_o$, $R_s$ and $R_l$ of $D^0$$D^0$ with respect to transverse pair momentum $K_T$ are shown.
The lines are results for $Y$ within the range of $(-1,1)$, while the symbols denote the results calculated in the LCMS.
In this paragraph, the results for $Y$ in the range of $(-1,1)$ are discussed first, followed by a discussion of the differences between these results and those obtained in the LCMS.
The squeezing effect leads to an increase in $R_o$, which is more pronounced at high $K_T$ values.
This is due to the impacts of the squeezing effect on the source distribution.
For $\beta = 0$, the squeezing effect does not affect the transverse source distribution.
The value of $R_s$ solely depends on the transverse spatial distribution of the source, so at $\beta = 0$,
the squeezing effect has no bearing on the value of $R_s$.
For $\beta > 0$, the squeezing effect leads to an increase in $R_s$, which is slightly more pronounced at high $K_T$ values.
The squeezing effect results in a significant increase in $R_l$.
This is due to the fact that the squeezing effect causes a significant widening of the source space-time rapidity distribution, leading to an expansion of the temporal distribution of and the longitudinal distribution of the source.
The squeezing effect has a slightly greater impact on $R_o$, $R_s$ and $R_l$ of $D^0$$D^0$ at $T=0.14$ GeV compared to $T=0.15$ GeV.
The value of $R_o$ calculated in the LCMS is slightly lower than that calculated for $Y$ within the range of $(-1,1)$, while
the value of $R_l$ calculated in the LCMS exceeds that calculated for $Y$ in the range of $(-1,1)$.
When considering the squeezing effect, this difference becomes more pronounced compared to when it is not taken into account.
In contrast, the value of $R_s$ calculated in the LCMS is identical to that calculated for $Y$ in the range of $(-1,1)$.
The influence of the squeezing effect on the HBT radii of $D^0$$D^0$ in the LCMS closely resembles its influence on the HBT radii of $D^0$$D^0$
for $Y$ in the range of $(-1,1)$.

\begin{figure}[htbp]
\vspace*{-1mm}
\includegraphics[scale=0.82]{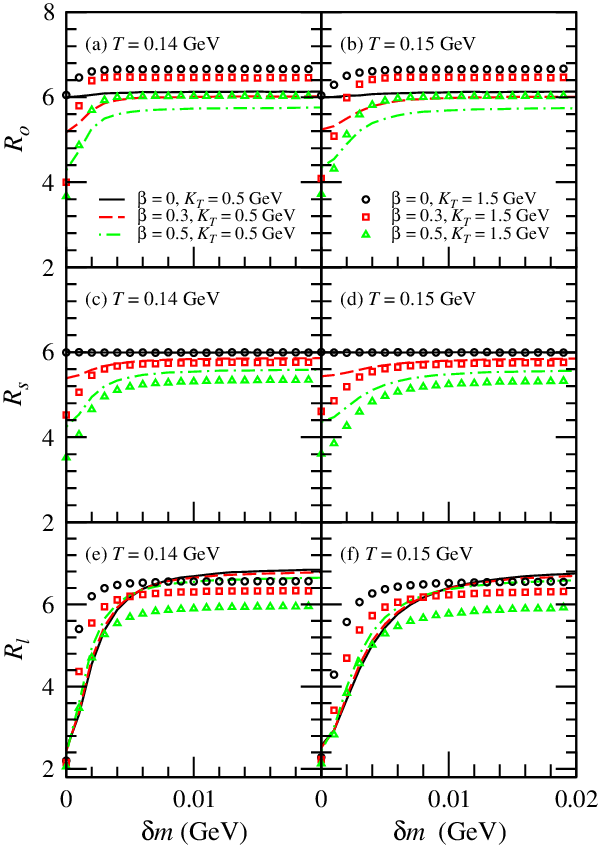}
\vspace*{0mm}
\caption{(Color online) HBT radii $R_o$, $R_s$ and $R_l$ of $D^0$$D^0$ with respect to $\delta m$ for $K_T$ = 0.5 GeV and 1.5 GeV. The plots illustrate the results of $R_o$ in (a) and (b), $R_s$ in (c) and (d), $R_l$ in (e) and (f). }
\label{roslddm}
\end{figure}

In Fig. \ref{roslddm}, the HBT radii $R_o$, $R_s$ and $R_l$ of $D^0$$D^0$ with respect to $\delta m$ for $K_T$ = 0.5 GeV and 1.5 GeV are presented.
When $\beta = 0$, the squeezing effect has no impact on $R_s$, which remains steady as $\delta m$ rises.
However, in other instances, as $\delta m$ increases, the value of $R_o$, $R_s$ and $R_l$ initially increase before leveling off, presenting a decreasing slope in the relationship between $R_o$, $R_s$ and $R_l$ and $\delta m$. Eventually, $R_o$, $R_s$ and $R_l$ stabilize with no changes as $\delta m$ further increases.

The squeezing effect exerts a greater influence on the values of $R_o$, $R_s$ and $R_l$ of $D^0$$D^0$ in comparison to $\phi$$\phi$,
because the mass of $D^0$ is greater than that of $\phi$.
The squeezing effect significantly influences the three-dimensional HBT radii, particularly $R_o$ and $R_l$, for cylinder expansion sources compared to the one-dimensional HBT radii for spherically symmetric sources. This is due to the notable impact of this effect on the source space-time rapidity distribution in cylinder expansion sources.

\subsection{$K^+K^+$}
\begin{figure}[htbp]
\vspace*{0mm}
\includegraphics[scale=0.66]{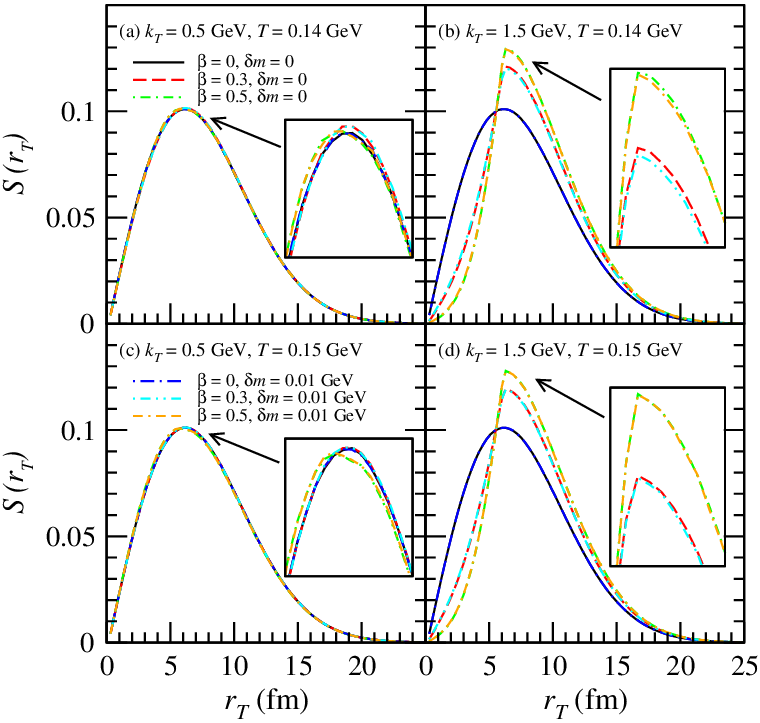}
\vspace*{0mm}
\caption{(Color online) Normalized distributions of transverse coordinate of the $K^+$ emission sources for $k_T$ = 0.5 GeV and 1.5 GeV.
Here, $T$ = 0.14 GeV in plots (a) and (b), while $T$ = 0.15 GeV in plots (c) and (d).  }
\label{psrk}
\end{figure}

\begin{figure}[htbp]
\vspace*{0mm}
\includegraphics[scale=0.66]{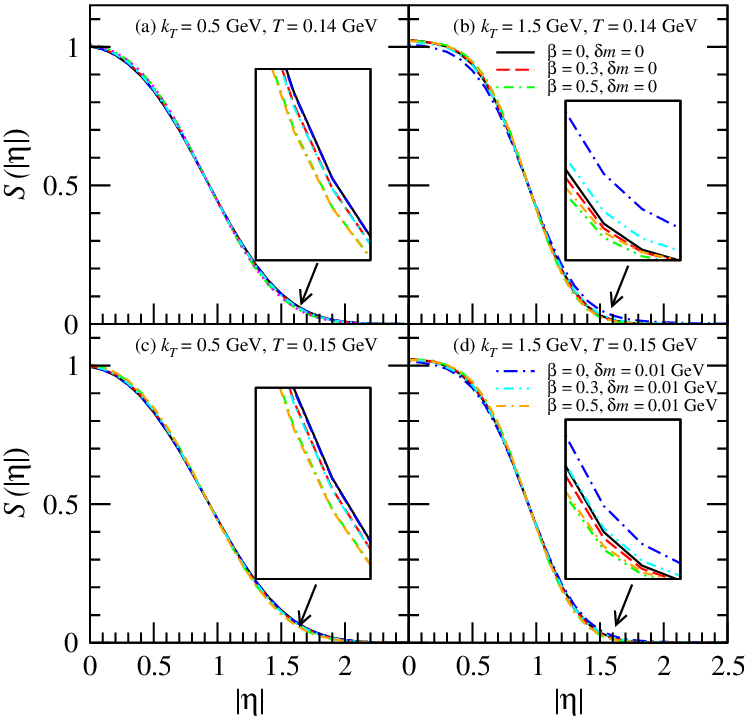}
\vspace*{0mm}
\caption{(Color online) Normalized distributions of $|\eta|$ of the $K^+$ emission sources for $k_T$ = 0.5 GeV and 1.5 GeV. }
\label{pszk}
\end{figure}

\begin{figure}[htbp]
\vspace*{0mm}
\includegraphics[scale=0.82]{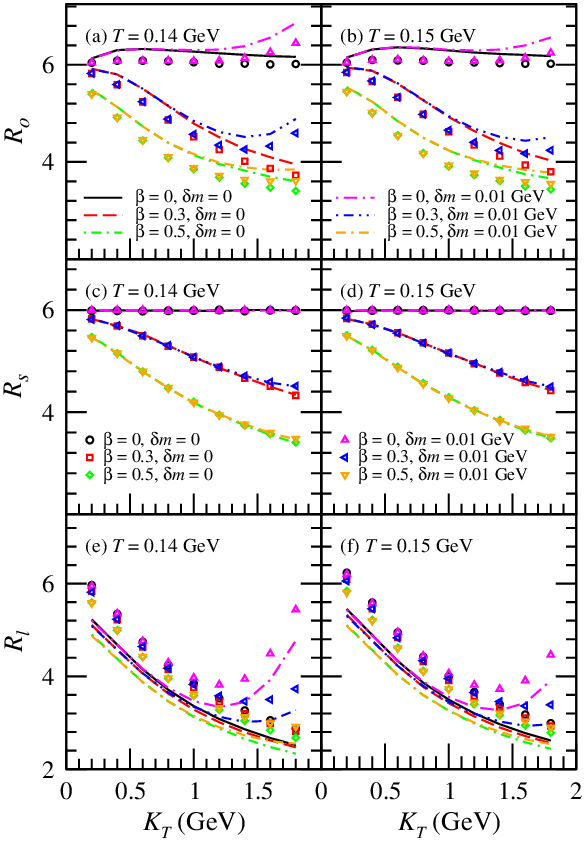}
\vspace*{0mm}
\caption{(Color online) HBT radii $R_o$, $R_s$ and $R_l$ of $K^+$$K^+$ with respect to transverse pair momentum $K_T$. The plots illustrate the results of $R_o$ in (a) and (b),
$R_s$ in (c) and (d), $R_l$ in (e) and (f). The lines are results for $Y$ within the range of $(-1,1)$, while the symbols denote the results calculated in the LCMS.}
\label{roslk}
\end{figure}

In Fig. \ref{psrk}, the normalized distributions of transverse coordinate of the $K^+$ emission sources for $k_T$ = 0.5 GeV and 1.5 GeV are presented.
Here, $T$ = 0.14 GeV in plots (a) and (b), while $T$ = 0.15 GeV in plots (c) and (d).
For $\beta = 0$, it indicates the absence of any velocity expansion in the transverse plane, and the transverse source distributions for $k_T$ = 1.5 GeV
being identical to those for $k_T$ = 0.5 GeV.
When the squeezing effect is not considered ($\delta m = 0 $), the transverse flow has a slight influence on the transverse source distributions at $k_T$ = 0.5 GeV.
For $\beta = 0.3$, the transverse flow induces a slight shift in these distributions toward larger transverse coordinates $r_T$ at $k_T$ = 0.5 GeV.
Conversely, at $\beta = 0.5$, the transverse flow causes a slight shift in the transverse source distributions towards smaller transverse coordinate $r_T$ for $k_T$ = 0.5 GeV.
This occurs because, at $\beta = 0.3$, the transverse flow within regions of small transverse coordinates $r_T$ is low, resulting in fewer production of $K^+$ particles with $k_T$ = 0.5 GeV.
The transverse flow induces a shift in the transverse source distributions toward larger transverse coordinates $r_T$ at $k_T$ = 1.5 GeV.
As the transverse flow increases, this phenomenon becomes more pronounced. At $T$ = 0.14 GeV, this phenomenon is slightly more significant than at $T$ = 0.15 GeV.
The mass of Kaon is expected to be reduced by around 0.01 GeV in the pionic medium \cite{pm}.
When considering the squeezing effect, the $\delta m$ for $K^+$ is taken as 0.01 GeV in Figs. \ref{psrk}-\ref{roslk}.
For $\beta = 0$, the squeezing effect does not impact the transverse source distribution.
However, for $\beta > 0$, the squeezing effect slightly decreases the influence of transverse flow on the transverse source distribution at $k_T$ = 1.5 GeV
while having no impact at $k_T$ = 0.5 GeV. This phenomenon is more pronounced for $T$ = 0.14 GeV.

In Fig. \ref{pszk}, the normalized distributions of $|\eta|$ of the $K^+$ emission sources for $k_T$ = 0.5 GeV and 1.5 GeV are shown.
For $\delta m = 0 $, the transverse flow has a slight reducing effect on the space-time rapidity distribution of the sources.
At $k_T$ = 0.5 GeV, the squeezing effect does not influence the source's space-time rapidity distribution.
However, at $k_T$ = 1.5 GeV, the squeezing effect causes a slight expansion of the source's space-time
rapidity distribution. This phenomenon is more pronounced with lower transverse flow and smaller freeze-out temperatures.

In Fig. \ref{roslk}, HBT radii $R_o$, $R_s$ and $R_l$ of $K^+$$K^+$ with respect to transverse pair momentum $K_T$ are shown.
Here, the lines are results for $Y$ within the range of $(-1,1)$, while the symbols denote the results calculated in the LCMS.
At high values of $K_T$, the squeezing effect results in a rise in both $R_o$ and $R_l$. The rise in $R_l$ is more significant than that of $R_o$.
This phenomenon is more obvious for larger $K_T$, as well as for lower transverse flow and freeze-out temperatures.
This is due to the influences of the squeezing effect on the source distribution.
For $\beta = 0$, the squeezing effect does not affect the transverse source distribution, so it also has no impacts on $R_s$.
For $\beta > 0$, the squeezing effect leads to a slight increase in $R_s$ at high $K_T$.
$R_s$ calculated in the LCMS is identical to that calculated for $Y$ in the range of $(-1,1)$.
$R_o$ calculated in the LCMS is slightly lower than that calculated for $Y$ in the range of $(-1,1)$, while $R_l$ calculated in the LCMS is greater than that calculated for $Y$ in the range of $(-1,1)$.
The squeezing effect's impacts on HBT radii of $K^+$$K^+$ in the LCMS are similar to the effect observed by $Y$ in the range of $(-1,1)$.
For a small in-medium mass modification,
the squeezing effect exerts a weaker influence on the HBT radii of $K^+$$K^+$ compared to $\phi$$\phi$ and $D^0$$D^0$, which can be attributed to the lower mass of $K^+$ relative to $\phi$ and $D^0$.

\begin{figure}[htbp]
\vspace*{-1mm}
\includegraphics[scale=0.82]{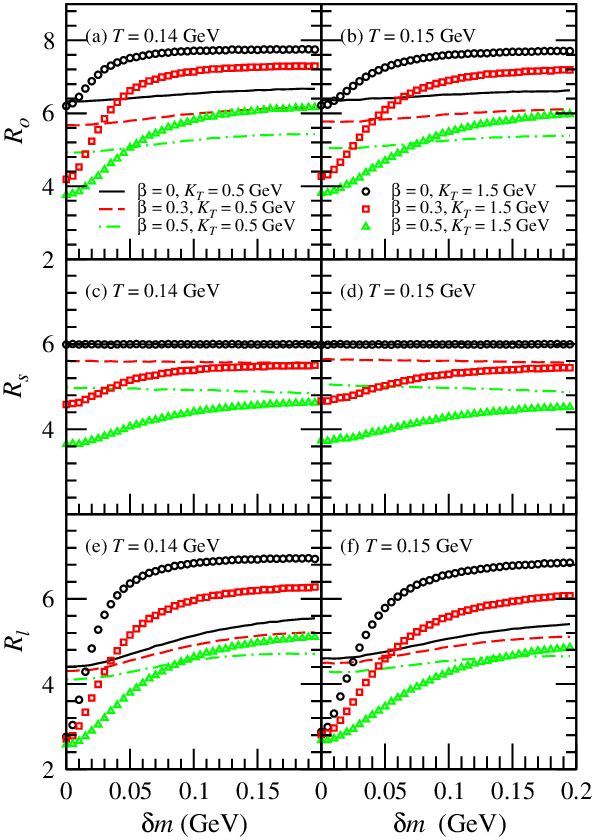}
\vspace*{0mm}
\caption{(Color online) HBT radii $R_o$, $R_s$ and $R_l$ of $K^+$$K^+$ with respect to $\delta m$ for $K_T$ = 0.5 GeV and 1.5 GeV. The plots illustrate the results of $R_o$ in (a) and (b), $R_s$ in (c) and (d), $R_l$ in (e) and (f). }
\label{roslkdm}
\end{figure}

In Fig. \ref{roslkdm}, the HBT radii $R_o$, $R_s$ and $R_l$ of $K^+$$K^+$ with respect to $\delta m$ for $K_T$ = 0.5 GeV and 1.5 GeV are shown.
For $\beta = 0$, the squeezing effect does not influence $R_s$, which remains constant as $\delta m$ increases.
For $\beta > 0$, as $\delta m$ increases, $R_s$ changes very little when $K_T$ = 0.5 GeV.
However, in other instances, as $\delta m$ increases, $R_o$, $R_s$ and $R_l$ first increase monotonically before saturating,
showing a decay in the slope of their correlation with $\delta m$. Ultimately, further increases in $\delta m$ result in no significant changes to
$R_o$, $R_s$ and $R_l$.

\section{Summary and discussion}
The interactions with the source medium result in a shift in the mass of bosons, inducing a squeezing effect.
In this paper, the impacts of the squeezing effect on the three-dimensional HBT radii are analyzed based on cylinder expansion sources.

The impacts of the squeezing effect on the three-dimensional HBT radii of $\phi$$\phi$, $D^0$$D^0$,
and $K^+$$K^+$ are analyzed.
The squeezing effect suppresses the impact of transverse flow on the transverse source distribution
and broadens the space-time rapidity distribution of the source, causing an increase in the HBT radii, particularly in out and longitudinal direction.
This phenomenon is more pronounced for higher transverse pair momentum $K_T$.
Consequently, the HBT radii no longer monotonically decrease with increasing $K_T$. This phenomenon is referred to the non-flow behavior of the HBT radii.
The squeezing effect has a slightly greater impact on $R_o$, $R_s$ and $R_l$ at $T=0.14$ GeV compared to $T=0.15$ GeV,
and it exerts a greater influence on the values of $R_o$, $R_s$ and $R_l$ of $D^0$$D^0$ in comparison to $\phi$$\phi$.
Additionally, the squeezing effect is also more significant for $\phi$$\phi$ than for $K^+$$K^+$.
The squeezing effect notably impacts the three-dimensional radii of the HBT, especially $R_o$ and $R_l$, in the case of cylinder expansion sources compared to the one-dimensional radii for spherically symmetric sources.

The non-flow behavior of $R_o$ and $R_l$ for $D^0$$D^0$ and $\phi$$\phi$ can be manifested in the small transverse momentum region compared to $K^+$$K^+$.
The non-flow behavior of $R_o$ and $R_l$ for $D^0$$D^0$ and $\phi$$\phi$ may be more likely to be measured experimentally compared to $K^+$$K^+$.
For small transverse flow, the squeezing effect has a more pronounced influence on $R_o$ and $R_l$ of $K^+$$K^+$ than it does for large transverse flow.
The non-flow behavior of $R_o$ and $R_l$ of $K^+$$K^+$ may be observed in the large transverse momentum region for small transverse flow.
However, the current transverse momentum range used for measuring HBT radii of $K^+$$K^+$ has not yet reached the region
where this non-flow behavior occurs \cite {khbt1,khbt2,khbt3,khbt4,khbt5,khbt6}. It should be emphasized that the effects demonstrated in this paper manifest only when particles undergo in-medium mass modifications. In the absence of such mass modifications, the non-flow behavior signals of the HBT radii will be entirely absent.

In this paper, the Gaussian form is used in the analysis of the three-dimensional HBT radii.
Due to the squeezing effect, the distribution of the source may actually differ from a Gaussian form.
Studying the effects of the squeezing effect on the source's shape is valuable, for instance, by employing a L$\acute{e}$vy-type form \cite{Tlevy,levy1,levy2,levy3}
to study how the squeezing effect influences both the three-dimensional HBT radii and the source's shape.

The cylinder expansion source used in this paper is more reasonable compared to the previously utilized spherically
symmetric source \cite{Zhang2024}. The independent transverse spatial distribution and temporal
distribution of the cylinder expansion source conflict with the actual situation.
Investigating the influences of the squeezing effect on three-dimensional HBT radii
through the utilization of more realistic sources, like hydrodynamical sources, presents an interesting direction for additional exploration.

\end{document}